# A simple demonstration of Bell's theorem involving two observers and no probabilities or inequalities

P.K.Aravind
Physics Department
Worcester Polytechnic Institute
Worcester, MA 01609
(email: paravind@wpi.edu)

ABSTRACT

This paper describes a device, consisting of a central source and two widely separated detectors with six switch settings each, that provides a simple *gedanken* demonstration of Bell's theorem without relying on either statistical effects or the occurrence of rare events. The mechanism underlying the operation of the device is revealed for readers with a knowledge of quantum mechanics.

## 1. Introduction

This paper presents a *gedanken* experiment involving a source and two widely separated detectors/observers that conveys the essence of Bell's theorem [1] to a lay audience without relying on either statistical effects or the occurrence of rare events. The device on which this demonstration is based was suggested by the work of David Mermin and Asher Peres, but the demonstration itself is due to me (see Sec.4 for a more detailed statement of credits).

The present demonstration is set within the same general framework as several of Mermin's earlier nontechnical demonstrations [2-4] of Bell's theorem. In Mermin's demonstrations, a central source emits several particles that fly off towards an equal number of widely separated detectors/observers. Each particle enters a detector, whose switch can be set to one of a small number of positions, and causes a light next to the chosen switch position to flash red or green. A complete demonstration with such a setup consists of a large number of repetitions of the following two basic steps: (1) a button is pressed on the source, releasing a bunch of particles that speed off towards their respective detectors; and (2) the observer manning each detector randomly sets its switch to one of the allowed positions and notes the color of the light that flashes when the particle enters his/her detector. At the end of all these runs the observers get together to compare their records of detector settings and light flashings. It is then that they discover that they have come face to face with the spookiness of quantum entanglement, which amounts to an informal appreciation of the central point of Bell's theorem.

Table 1 lists the salient features of Mermin's three earlier demonstrations of Bell's theorem, with the corresponding features of the present scheme listed underneath. One conspicuous difference between the present scheme and the earlier ones is that two particles now go to each detector, rather than just one. However the more significant differences are the numbers in the third and fourth columns. The present scheme involves only two detectors (an advantage over the GHZ-Mermin scheme) but each detector now has six switch settings (a disadvantage compared to all the other schemes). An advantage of the present scheme over the Bell-Mermin scheme is that it does not rely on statistical features of the data to produce its effects, while an advantage over the Hardy-Mermin scheme is that it does not rely on the occurrence of rare events. However a disadvantage vis-a-vis the earlier schemes is that the technology for implementing the present scheme in the laboratory is more complex and has not yet been fully developed.

| Scheme | # Particles | # Detectors or Observers | # Detector settings |
|---|---|---|---|
| Bell-Mermin [1,2] | 2 | 2 | 3 |
| GHZ-Mermin [3,5] | 3 | 3 | 2 |
| Hardy-Mermin [4,6] | 2 | 2 | 2 |
| Present scheme | 4 | 2 | 6 |

Table 1. Salient features of several nontechnical demonstrations of Bell's theorem.

## 2. The *gedanken* experiment

A source S emits four particles, two of which fly off to the left towards Alice and the other two to the right towards Bob (see Fig.1). Each pair of particles enters a detector which performs a measurement on it and displays the results on a screen segmented into nine square panels arranged in the form of a 3 x 3 array, as shown in Fig.1. Alice and Bob can each set a switch on their detector to one of six positions, thereby causing an entire row or column of panels on it to light up in response to the incoming particles. Each panel that lights up upon receipt of the particles lights up either red or green (the journal is unfortunately unable to print in color, so red will appear everywhere in the figures as black and green as grey, much to my regret!). The six switch settings on each detector will be denoted R1, R2 and R3 (for the three rows of panels, from top to bottom) and C1, C2 and C3 (for the three columns of panels, from left to right). Figure 1 shows a run with this setup in which Alice chooses the detector setting R1 and Bob the setting C2, and their panels light up as shown.

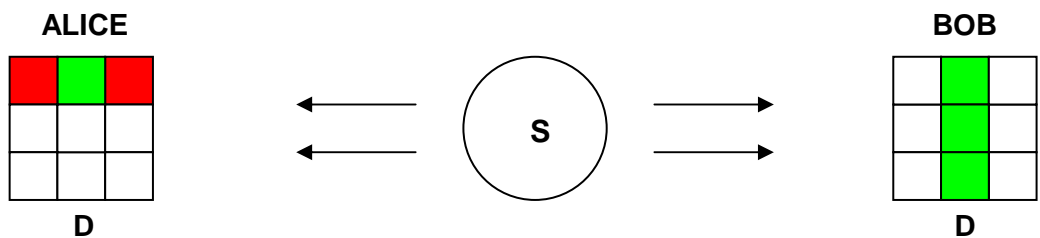

Fig.1. The *gedanken* experiment. A source **S** emits four particles, two of which move off to the left and the other two to the right. Each pair of particles enters a detector **D**, adjusted to one of six possible settings, and causes certain panels on it to light up in the manner described in the text. In the run above, Alice chooses the setting R1 on her detector and causes the panels in the first row to light up red, green, red (from left to right), while Bob chooses the setting C2 and causes all the panels in the second column to light up green.

A complete demonstration with the above device consists of a large number of repetitions of the following two basic steps: (1) a button is pressed on the source, releasing four particles, two of which proceed towards Alice and the other two towards Bob, and (2) Alice and Bob each independently and randomly set their detector to one of its six settings and note the colors of the panels that light up upon entry of the particles. After a large number of such runs, Alice and Bob get together to compare their records of detector settings and light flashings. When they do this they find that all their observations, without exception, can be summarized in the form of two simple rules, which we now state.

**Rule 1 (the "parity" rule):** For any of the detector settings R1, R2, R3, C1 or C2, an even number of panels lights up red and an odd number lights up green. However, for the setting C3, an odd number of panels lights up red and an even number lights up green. Further, the four possible outcomes for each detector setting occur randomly (i.e. with a probability of ¼ each).

Figure 2 illustrates this rule by showing the four ways in which the panels can light up for each of the six detector settings.

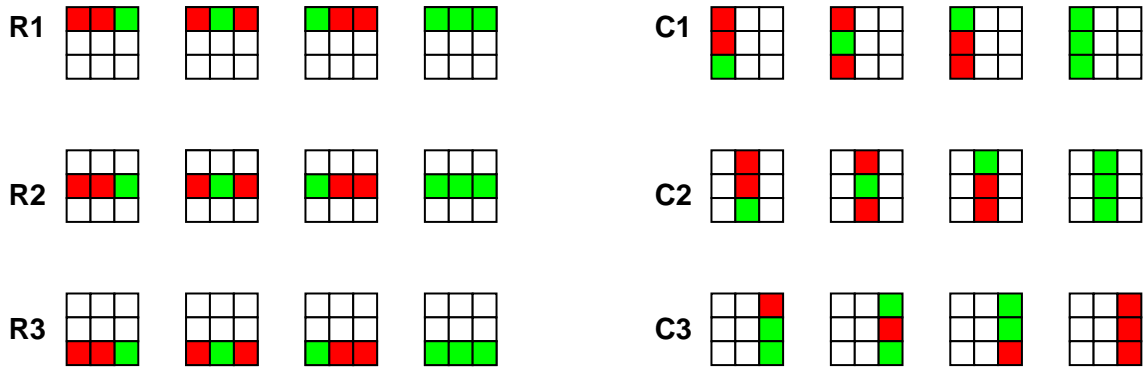

Fig.2. Illustrating Rule 1 (the "parity" rule)

**Rule 2 (the "correlation" rule):** In those runs in which Alice's and Bob's detector settings cause one or more common (i.e. similarly placed) panels on their detectors to light up, the common panels always light up the same colors.

This rule is illustrated in Figure 3, which shows Alice and Bob's detector responses alongside each other for a number of runs. If one looks at the first, second and fourth of the runs listed, one sees that all the common panels that light up on both detectors always have the same colors. However no common panels light up in the third run, and Rule 2 does not apply in this case. Note that Rule 1 is always obeyed by both detectors in all the runs.

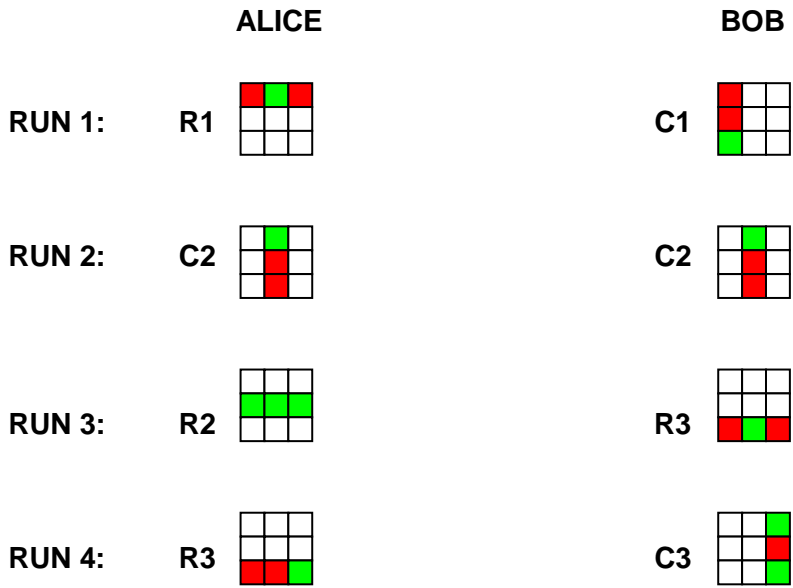

Fig.3. Illustrating Rules 1 and 2 in a series of runs carried out by Alice and Bob.

The above rules present us with an interesting puzzle: how can the source and detectors be constructed so as to act in conformity with these rules at all times?

A useful first step in tackling this puzzle is to realize that the properties of the source particles, as revealed by the colors of the detector panels, are "elements of reality" in the sense of Einstein, Podolsky and Rosen [7]. What this means is simply that these properties can be determined without disturbing the particles, or the detectors they interact with, in any way. Suppose, for example, that one wishes to determine the property of Bob's source particles revealed by the panel in the first row and second column of his detector. One can do this by having Alice use either the setting R1 or C2 and observe the color of the same panel on her detector; one can then predict, from the infallibility of Rule 2, that Bob's panel must also have that same color. (It should be mentioned that, in the above demonstration, matters are arranged so that Alice's and Bob's detectors always fire within a very short interval of each other, leaving insufficient time for a light signal to be transmitted from one to the other; this ensures that there can be no disturbance of the particles/detector on one side as a result of any actions affecting the particles/detector on the other). It should be stressed that all nine properties of Bob's particles, as revealed by the colors of his detector panels, must be elements of reality in each run, because there is no telling which property Bob could force to reveal itself through his choice of detector setting (and the revealed property must, of course, always agree with Alice's prediction of it).

The above reasoning leads to the conclusion that, in any run, each of Bob's detector panels must have a perfectly predetermined color. And, by exchanging the roles of Alice and Bob in the argument, one can conclude the same about Alice's panels as well. (In fact, one can go further and conclude, from the correlation rule, that the colors of Alice's and Bob's panels must match each other perfectly in every run).

The solution to our puzzle thus reduces to the task of finding solutions to the following problem: assign a definite color, red or green, to each of the nine detector panels in such a way that all the constraints imposed by Rule 1 (the "parity" rule) are met. However this task is easily seen to be impossible by asking about the total number of red panels on a detector: on the one hand, Rule 1 requires this number to be even (if one sums the red panels over the rows) but, on the other, it also requires this number to be odd (if one sums the red panels over the columns). This contradiction shows that there is no logical solution to our puzzle, which therefore begins to take on the air of a magic trick. What, not so fast, you say? Okay, let's listen to your explanation …

Maybe you cracked the puzzle, or maybe you didn't. If you'd like to read my explanation, here it is.

### 3. How the trick is done

When the button is pressed on the source, it emits four spin-1/2 particles ("qubits") in the state

$$|\Psi\rangle = \frac{1}{\sqrt{2}}\left(|0\rangle_1|0\rangle_2 + |1\rangle_1|1\rangle_2\right) \otimes \frac{1}{\sqrt{2}}\left(|0\rangle_3|0\rangle_4 + |1\rangle_3|1\rangle_4\right), \qquad (1)$$

where $|0\rangle_i$ and $|1\rangle_i$ denote a pair of orthonormal basis states of qubit $i$ ($i = 1,..,4$). Qubits 1 and 3 of this state go to Alice, while qubits 2 and 4 go to Bob. In other words, the source emits a pair of Bell states, with one member of each Bell state going to Alice and the other members to Bob.

It should be added that the states $|0\rangle_i$ and $|1\rangle_i$ are eigenstates, with eigenvalues +1 and -1 respectively, of the Pauli operator $\sigma_z$ of qubit $i$ (these states are often referred to now as the "standard" or "computational" basis states).

Figure 4 shows nine observables for a pair of qubits arranged in the form of a 3 x 3 array, with the observables in each row or column of this array forming a mutually commuting set. Each observable has only the eigenvalues ±1 and, further, the product of the observables in any row or column is $+I$, with the exception of the last column for which the product is $-I$ ($I$ being the identity operator). When any of the detector settings R1,R2 … or C3 is chosen, the detector carries out a measurement of the commuting observables in the corresponding row or column of Fig.4 on the two qubits entering it and depicts the observed eigenvalues as colored lights on its panels according to the convention that a +1 is a green and a -1 a red. Rule 1 then follows as an immediate consequence of the fact [8] that if several mutually commuting observables obey a certain functional relationship, their measured eigenvalues in an arbitrary state also obey that relationship; in the present context, this implies that the product of the measured eigenvalues of the observables in each row or column of Fig.4 must be +1, with the exception of the last column for which the product must be -1. The last statement, when translated into the language of the red and green lights, is nothing but the parity rule.

| $1\otimes\sigma_z$ | $\sigma_z\otimes 1$ | $\sigma_z\otimes\sigma_z$ |
|---|---|---|
| $\sigma_x\otimes 1$ | $1\otimes\sigma_x$ | $\sigma_x\otimes\sigma_x$ |
| $\sigma_x\otimes\sigma_z$ | $\sigma_z\otimes\sigma_x$ | $\sigma_y\otimes\sigma_y$ |

Fig.4. The Mermin-Peres "magic square" [8,9]. Each entry in the square is an observable for a pair of qubits, with $1, \sigma_x, \sigma_y$ and $\sigma_z$ denoting the identity and Pauli operators of a qubit. The observables in each row or column of the square form a mutually commuting set. When a switch setting on a detector is chosen, it carries out a measurement of the commuting observables in one of the rows or columns above. The first and second operators in each observable refer to qubits 1 and 3 for Alice's detector and to qubits 2 and 4 for Bob's detector.

The origin of Rule 2 can be understood as follows. Let $|\psi_i\rangle$ $(i=1,..,4)$ be an arbitrary set of orthonormal states in the joint space of qubits 1 and 3 and suppose that they have the expansion $|\psi_i\rangle = a_i|0\rangle_1|0\rangle_3 + b_i|0\rangle_1|1\rangle_3 + c_i|1\rangle_1|0\rangle_3 + d_i|1\rangle_1|1\rangle_3$ where the $a_i,...,d_i$ are suitable

complex coefficients. Then $|\phi_i\rangle = a_i^*|0\rangle_2|0\rangle_4 + b_i^*|0\rangle_2|1\rangle_4 + c_i^*|1\rangle_2|0\rangle_4 + d_i^*|1\rangle_2|1\rangle_4$ $(i=1,..,4)$ are evidently an orthonormal set of states in the joint space of qubits 2 and 4. It can be verified that the state (1) can be rewritten in terms of the $|\psi_i\rangle$ and $|\phi_i\rangle$ as

$$|\Psi\rangle = \frac{1}{2}\left[|\psi_1\rangle|\phi_1\rangle + |\psi_2\rangle|\phi_2\rangle + |\psi_3\rangle|\phi_3\rangle + |\psi_4\rangle|\phi_4\rangle\right] . \qquad (2)$$

When Alice carries out a measurement of one of the sets of commuting observables in Fig.4, she projects her qubits into one of the eigenstates $|\psi_i\rangle$ of this set and Bob's qubits into the associated state $|\phi_i\rangle$. However it turns out that the coefficients $a_i,...,d_i$ are always real for all of the eigenstates arising from Fig.4, and hence that each $|\phi_i\rangle$ has the same form as the corresponding $|\psi_i\rangle$ (when expressed in terms of its own standard basis). It therefore follows that if Bob measures one or more of the same observables as Alice, he always finds the same eigenvalues as her for them, which is just the correlation rule. The expression (2) also explains the fact, mentioned at the end of Rule 1, that all four outcomes for each detector setting occur with the same probability (of ¼).

### 4. Credits for the demonstration

The "magic square" of Fig.4, which lies at the heart of the present demonstration, is due to Mermin [8], who built on an earlier suggestion of Peres [9]. Mermin [8,10] used this array of observables to prove the Bell-Kochen-Specker(BKS) theorem [11], a close relative of the more famous Bell's theorem. Peres [12] also used this array to give a related, but different, proof of the BKS theorem. The discovery that the Mermin-Peres proof of the BKS theorem could be converted into a proof of Bell's theorem is more recent. Cabello [13] and the author [14] showed, in slightly different ways, how this could be done by distributing one member each of a pair of Bell states to two observers and having them carry out certain measurements on their particles. It is the author's version of this proof of Bell's theorem "without inequalities" that has been turned into the present non-technical demonstration. This very brief survey of the literature makes no attempt at completeness and merely tries to highlight the works that directly influenced this demonstration.

After an earlier version of this paper had been posted on the eprint archive, Richard Cleve informed me that David Mermin and he had come up with a similar scheme in which each detector had only three switch settings. This is easily accomplished in the present framework by allowing Alice to use only the row settings R1, R2 and R3 on her detector and Bob to use only the column settings C1,C2 and C3 on his. It is not difficult to see that this restriction still leads to the same magical effect as before. Rule 1 can also be cast in a more symmetrical form, if desired, by affixing a negative sign to the second and third observables in the last row of Fig.4; then the parity of the rows (or columns) with respect to the red squares is always even (or odd).